PLANETARY SCIENCE

# Simultaneous Mars-orbit observations reveal Kelvin-Helmholtz instability–driven bulk atmospheric ion escape

**Chi Zhang[1]*, Chuanfei Dong[1,2]*, Gangkai Poh[3,4], Jasper Halekas[5], Xuanye Ma[6], Ruhunusiri Suranga[7], Kathleen G. Hanley[8], Han-Wen Shen[5], Hongyang Zhou[1], Xinmin Li[1], Liang Wang[1], Jiawei Gao[1], Shannon Curry[7], Christian Mazelle[9]**

**Atmospheric ion escape driven by the solar wind is a key process controlling the long-term loss of the Martian atmosphere. Localized plasma clouds can carry substantial fluxes of planetary ions away from Mars, representing episodes of bulk escape. However, their origin has remained unclear due to the absence of simultaneous upstream measurements. Using joint observations from the MAVEN and Tianwen-1 missions, which provide real-time upstream monitoring, we present direct evidence that these plasma clouds are nonlinear wave packets generated by the Kelvin-Helmholtz instability (KHI). The spatial scale of KH waves is constrained for the first time via two-point measurements. Ion fluxes within plasma clouds are one to two orders of magnitude higher than those in typical steady-state escape channels. Our results indicate that KHI is an important process for solar wind coupling to planetary upper atmospheres and plays a crucial role in shaping atmospheric ion escape for unmagnetized planets.**

## INTRODUCTION

Multiple lines of evidence indicate that, billions of years ago, Mars had a global dynamo magnetic field, a much thicker atmosphere, and stable surface liquid water, broadly comparable to conditions on modern Earth [e.g., (*1*–*3*)]. Today, however, Mars lacks a global dipole field and surface liquid water and retains only a tenuous, $CO_2$ (carbon dioxide)-dominated atmosphere, implying substantial atmospheric loss over time. Understanding how Mars's atmosphere eroded and evolved is central to reconstructing its climate history and assessing past habitability (*4*–*6*).

At present, the total atmospheric oxygen escape rate from Mars is $\sim 6 \times 10^{25}\ s^{-1}$ and is dominated by photochemical neutral escape, primarily driven by hot oxygen atoms produced through dissociative recombination (*4*, *7*–*10*). This process contributes $\sim 5 \times 10^{25}\ s^{-1}$, accounting for the majority of the total oxygen loss. The second major escape channel is ion escape, in which the solar wind (a fast stream of charged particles from the Sun) directly interacts with the upper atmosphere, transfers energy to atmospheric ions, and accelerates them to velocities sufficient to overcome gravity and escape into interplanetary space (*11*–*17*). However, although ion escape represents a smaller fraction of the present-day atmospheric loss, it is likely to have played a much more substantial role over geological timescales. Under the more intense solar wind conditions of early Mars, ion escape would have been substantially enhanced and may have even exceeded photochemical escape as the dominant loss mechanism (*5*).

[1]Center for Space Physics and Department of Astronomy, Boston University, Boston, MA, USA. [2]School of Natural Sciences, Institute for Advanced Study, Princeton, NJ, USA. [3]Center for Research and Exploration in Space Sciences and Technology II, Catholic University of America, Washington, DC, USA. [4]Solar System Exploration Division, NASA Goddard Space Flight Center, Greenbelt, MD, USA. [5]Department of Physics and Astronomy, University of Iowa, Iowa City, IA, USA. [6]Embry-Riddle Aeronautical University, Daytona Beach, FL, USA. [7]Laboratory for Atmospheric and Space Physics, University of Colorado, Boulder, CO, USA. [8]Space Sciences Laboratory, University of California, Berkeley, Berkeley, CA, USA. [9]Institut de Recherche en Astrophysique et Planétologie, CNRS, Université de Toulouse, CNES, Observatoire Midi-Pyrénées, Toulouse, France.
*Corresponding author. Email: zc199508@bu.edu (C.Z.); dcfy@bu.edu (C.D.)

It is well established that Martian atmospheric ion escape proceeds primarily via two quasi-steady channels. The first is the dayside ion plume, which originates in the topside ionosphere and extends beyond the bow shock; it comprises ionospheric ions accelerated by the solar-wind motional (convective) electric field ($\vec{E}_{sw}$) (*18*). The second is nightside magnetotail escape, in which a substantial fraction of ionospheric ions are transported into the magnetotail region and ultimately lost to interplanetary space (*19*–*21*).

Beyond the above two steady pathways, ion loss can also occur transiently through localized processes that produce brief episodes of bulk escape. A representative example is the plasma cloud, characterized by short-lived enhancements in escaping planetary ions (*22*, *23*). The origin of these clouds remains debated, with several candidate mechanisms proposed. Russell *et al.* (*24*) proposed that they are fundamentally associated with current sheet structures, whereas Penz *et al.* (*25*) suggested that the Kelvin-Helmholtz instability (KHI) could give rise to these clouds. Halekas *et al.* (*26*) later identified periodic plasma clouds near the ion composition boundary (ICB), which is the interface between solar wind and planetary ions enhancements. They proposed that planetary ions within these clouds are accelerated by magnetic tension forces, in a mechanism akin to a"snowplow" effect. Subsequent studies further linked plasma clouds to ionospheric irregularities (*27*) or to solar wind compressional structures (*28*). More recent observational studies (*29*–*32*) further supports the KHI-driven origin initially proposed by Penz *et al.* (*25*).

To elucidate the mechanisms underlying plasma clouds, it is essential to characterize the upstream solar wind conditions as well as the background electromagnetic fields and plasma flows in which they develop. Unlike Earth, which has an intrinsic dipole magnetic field, Mars hosts an induced magnetosphere generated by the interaction between the solar wind and its upper atmosphere (*33*–*36*). As a result, the electromagnetic and plasma environment at Mars is highly sensitive to upstream solar wind conditions, leading to stronger variability compared to intrinsic magnetospheres. However, previous observational studies have relied solely on single-spacecraft measurements without simultaneous upstream solar wind monitoring.

Consequently, it has often been assumed that the solar wind remains quasi-steady over timescales longer than a spacecraft's orbital period (i.e., several hours). This assumption introduces notable uncertainties, as the solar wind is inherently dynamic, particularly with respect to the interplanetary magnetic field (IMF) (*37*, *38*). The lack of real-time upstream context has therefore hindered efforts to accurately determine the properties of plasma clouds and to identify the physical processes responsible for their formation.

The arrival of the Tianwen-1 mission in 2021 (*39*), operating in conjunction with the MAVEN mission (*40*), established a dual-spacecraft observation system that helps overcome this limitation. First, when Tianwen-1 is positioned in the upstream solar wind, it provides real-time measurements of both the solar wind and its embedded IMF, enabling precise characterization of plasma cloud velocities and field properties relative to the background environment. Second, it enables us to consider whether the observed plasma clouds are likely driven by solar wind disturbances or can be intrinsically generated under steady-state solar wind conditions. Third, simultaneous two-point observations offer direct constraints on the spatial scale of plasma clouds: If both spacecraft reliably detect the same event, then its size must exceed their separation; if only one detects it, then the cloud must be smaller than the interspacecraft spacing.

Leveraging these capabilities, we conduct an investigation of plasma clouds using combined Tianwen-1 and MAVEN measurements. Our results show that the observed signatures of plasma clouds are most consistent with nonlinear wave packets generated by the KHI. These findings indicate that KHI plays a role in regulating atmospheric ion escape at Mars and is an important process in governing solar wind interactions with planetary upper atmosphere.

## RESULTS

### Representative cases of plasma clouds

Figure 1 presents a representative plasma cloud event observed between 04:25 and 05:00 UT on 31 July 2023. During this interval, MAVEN was located at an average position of (−0.72, −1.15, −1.25) $R_M$ (Mars radius, $R_M$= 3390 km) in the Mars solar orbital (MSO) coordinates (see Materials and Methods), approaching the nominal magnetic pile-up boundary (MPB) (Fig. 1, A and B). Although the MPB and the ICB are defined differently, they are generally located in close proximity (*41*). Therefore, we adopt the nominal MPB position from Trotignon *et al.* (*41*) as a proxy for the ICB and refer to it as the ICB throughout the paper.

Before 04:34 UT, MAVEN remained in the magnetosheath, detecting only protons with energies of 50 to 500 eV (Fig. 1C) with no clear signature of planetary heavy ions (Fig. 1, D and E). After 04:52 UT, the spacecraft entered the induced magnetosphere, where mainly planetary heavy ions were observed.

The interval of interest is 04:34 to 04:52 UT (marked by the gray shaded interval in Fig. 1, C to H), during which MAVEN repeatedly observed alternating populations of magnetosheath protons and planetary heavy ions ($O^+$ and $O_2^+$, see the black vertical dashed lines in Fig. 1, C to F). These signatures are consistent with the plasma cloud phenomena previously reported by Halekas *et al.* (*26*). The recurrence period is ∼2 min, and we therefore classify this event as a case of quasi-periodic plasma clouds. These clouds are characterized by strong magnetic field fluctuations, most notably manifested as impulsive spikes during the spacecraft's exit from the cloud (Fig. 1F). Throughout the entire cloud interval, Tianwen-1 was located in the upstream solar wind (see Fig. 1A), where it detected strong wave activity (Fig. 1G). Aside from these waves, the background IMF remained generally steady, with an average value of (−3.12, 2.94, −0.13) nT. This also excludes the possibility that the plasma clouds originated from upstream perturbations; instead, they most likely resulted from the interaction between the steady solar wind and the Martian ionosphere.

The clock angle, ϕ, is defined as the angle between the projected magnetic fields and $\vec{Z}_{MSO}$ in the $\vec{Y}_{MSO} - \vec{Z}_{MSO}$ plane, with the angle rotationally increasing from $+\vec{Z}_{MSO}$ toward $+\vec{Y}_{MSO}$. Hence, $\phi = 90°$ (or 270°) indicates that the projected magnetic field points toward $+\vec{Y}_{MSO}$ (or $-\vec{Y}_{MSO}$). In an ideal induced magnetosphere, the clock angle would remain unchanged as the IMF propagates toward Mars and becomes draped. As shown in Fig. 1H, the magnetic field clock angle in the magnetosheath differed by ∼30° from the solar wind values measured by Tianwen-1. This is consistent with the characteristic differences between solar wind and magnetosheath clock angles reported by Dong *et al.* (*42*) and Cheng *et al.* (*43*). However, during the cloud interval, the magnetic fluctuations produced pronounced variations in the clock angle, indicating that the plasma clouds were associated with notable distortions of the induced magnetic field.

Figure 2 shows a representative case of an isolated plasma cloud observed between 00:52 and 01:03 UT on 19 October 2022. During this interval, MAVEN was also located near the nominal ICB position, while Tianwen-1 remained in the upstream solar wind (Fig. 2, A and B). Before 00:55:40 UT and after 00:58:00 UT, MAVEN primarily detected magnetosheath protons (Fig. 2C). In contrast, between 00:55:40 UT and 00:58:00 UT (marked by the gray shadow vertical dashed lines in Fig. 2, C to H), we observe the enhancements in $O^+$ and $O_2^+$ fluxes (Fig. 2, D and E). As the spacecraft exited this region of enhanced flux, it also detected impulsive magnetic field spikes (Fig. 2F). Together, these features signal the passage of a plasma cloud.

However, unlike the quasi-periodic event, this plasma cloud appears as a single, isolated structure. A weaker signature is present near 00:59 UT, but it is much less distinct. Throughout the cloud interval, Tianwen-1 observed relatively steady magnetic fields (Fig. 2G), with an average value of (−0.026, −2.63, −0.12) nT. As shown in Fig. 2H, MAVEN and Tianwen-1 recorded similar magnetic field clock angles for most of the interval, except during the cloud passage.

### Properties of the plasma clouds

To investigate the properties of the observed clouds in detail, we construct the Mars Solar Electric coordinate system (see Materials and Methods) based on the average magnetic field measured by Tianwen-1. As shown in fig. S1, both isolated and quasi-periodic events were observed in the −E hemisphere, where $\vec{E}_{SW}$ points toward the planet. Figure 3 (A1 to F1) present a 10-min interval extracted from the quasi-periodic event shown in Fig. 1, containing three clouds. In contrast, Fig. 3 (A2 to F2) highlight the isolated cloud event. Given that the plasma clouds were moving tailward at high speeds relative to MAVEN, the spacecraft intercepted each cloud from the downstream edge, traversed through the cloud core, and exited at the upstream edge. Accordingly, we define the left side of each cloud as the downstream edge and the right side as the upstream edge. As illustrated in Fig. 3, both quasi-periodic and isolated events exhibit several common signatures in Mars solar electric (MSE) coordinates.

First, the magnetic field and density variations differ between the two edges. From the downstream edge toward the cloud center, the $B_y^{MSE}$ component and total field strength ($B_t$) gradually decrease (see Fig. 3, A1 and A2), with the increase of heavy ions density ratio (see Fig. 3, C1 and C2), indicating that the downstream edge is a gradual transition from magnetosheath protons to planetary heavy ions. In contrast, the upstream edge displays a sharp structure: the $B_Z^{MSE}$ component first increases positively and then rapidly reverses to negative, forming a clear $+B_Z^{MSE}$ to $-B_Z^{MSE}$ bipolar signature. Simultaneously, both $B_y^{MSE}$ and $B_t$ increase rapidly, forming an impulsive spikes pattern. These features, combined with an increase in total ion density (see Fig. 3, B1 and B2), suggest that the upstream edge acts as a sharp and compressional boundary. The minimum variance analysis hodogram of the upstream edge is not circular (see fig. S2), suggesting that it is unlikely to be a magnetic flux rope.

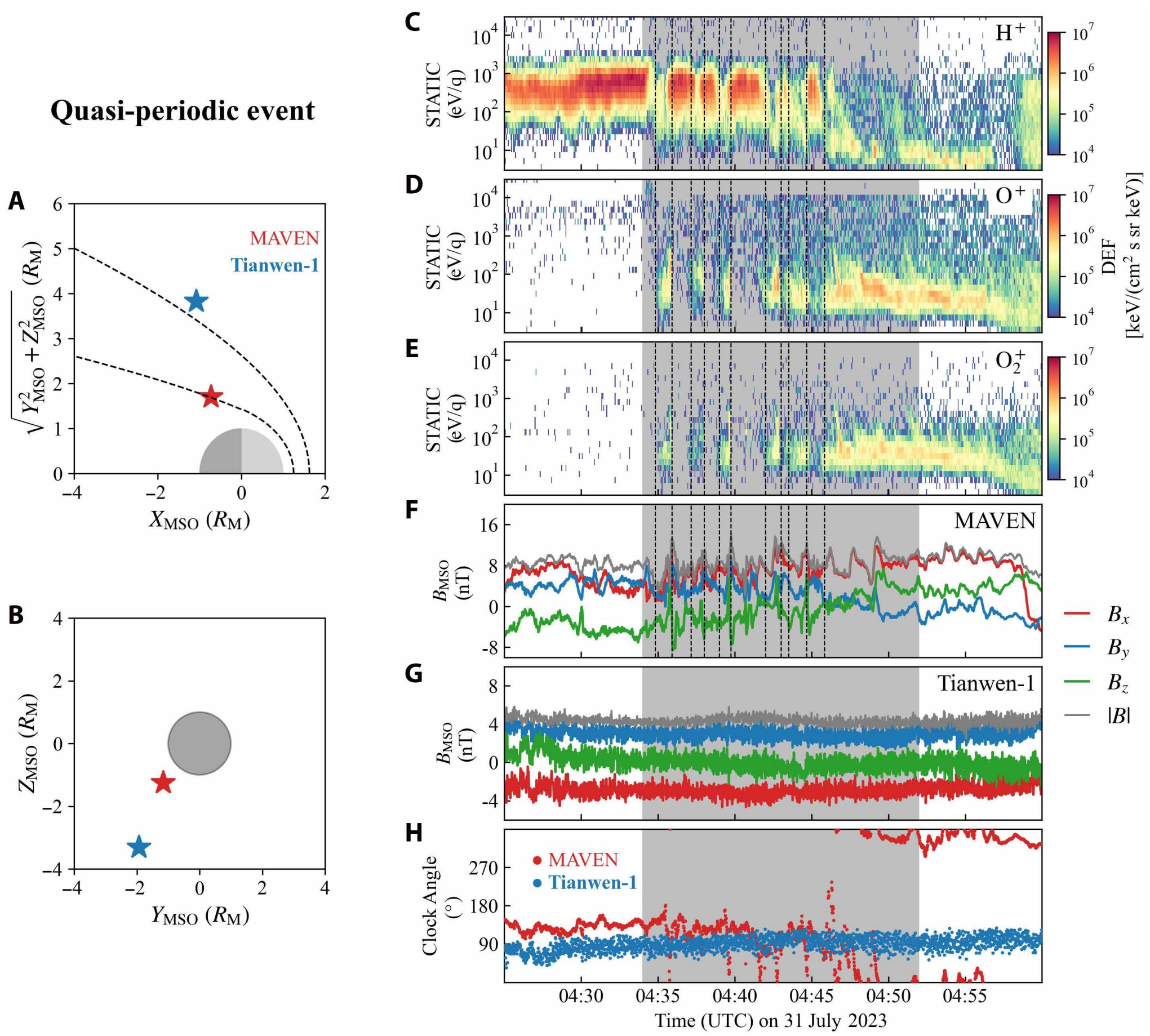

**Fig. 1. Overview of a representative quasi-periodic plasma cloud event observed on 31 July 2023.** (**A**) Average spacecraft locations in the $X_{MSO} - \sqrt{Y_{MSO}^2 + Z_{MSO}^2}$ plane. The red star and blue star denote MAVEN and Tianwen-1, respectively. The dashed curves in (A) indicate the nominal bow shock (BS) and magnetic pileup boundary (MPB). (**B**) Average spacecraft locations in the $YZ_{MSO}$ plane; the gray circle represents Mars. (**C**) $H^+$ energy spectra. (**D**) $O^+$ energy spectra. (**E**) $O_2^+$ energy spectra. (**F**) Magnetic fields observed by MAVEN in MSO coordinates. (**G**) Magnetic fields observed by Tianwen-1 in MSO coordinates. (**H**) Magnetic field clock angles. The gray shaded interval marks the time interval of the quasi-periodic plasma clouds. The black vertical dashed lines mark transitions between magnetosheath protons and planetary heavy ions.

Second, the proton bulk velocity exhibits large variations across the cloud edges. As the spacecraft enters the cloud in the downstream edge, the magnetosheath negative proton $V_x^{MSE}$ decreases (Fig. 3, D1 and D2), whereas it increases markedly at the upstream edge as the spacecraft exits the cloud. In addition, systematic changes are also observed in the $V_z^{MSE}$ component. At the downstream edge, the proton velocity shows $\delta V_z^{MSE} > 0$ (see the red arrows in Fig. 3, E1 and E2), while at the upstream edge, $\delta V_z^{MSE} < 0$ (see the blue arrows in Fig. 3, E1 and E2). Here, $\delta V_z^{MSE}$ denotes the change in the $V_z^{MSE}$. To assess whether the observed variation in $\delta V_z^{MSE}$ could be an artifact caused by the instrument's limited field of view (FOV), we examined the angular distribution of protons (see fig. S3). The results indicate that most of the protons are within the FOV, suggesting that the velocity variations are real rather than due to instrumental effects.

Last, as shown in Fig. 3 (F1 and F2), the total pressure (sum of magnetic and plasma thermal pressure) decreases from the downstream edge toward the cloud center and then sharply increases at the upstream edge, forming a pronounced local minimum at the cloud center and a local peak at the upstream edge.

These cloud features are consistent with Kelvin-Helmholtz (KH) wave packet forming at the interface between magnetosheath protons and planetary heavy ions. Figure 4 presents a simplified sketch of a single KH wave packet for illustrative purposes, although in reality, multiple wave packets may exist. Because KH waves primarily propagate tailward, MAVEN moves sunward relative to the wavefront (as indicated by the pink dashed arrow). As the spacecraft traverses the bulge-like KH wave packet, it observes vortex-like proton flows along with an enhanced density ratio of planetary heavy ions. Specifically, at the downstream edge of a KH wave, it encounters upward-directed magnetosheath protons, producing $\Delta V_z^{MSE} > 0$. Conversely, at the upstream edge, the flow is directed downward, resulting in $\Delta V_z^{MSE} < 0$. These vortex flows can also induce magnetic field rotations, which in turn lead to the observed variations in $B_Z^{MSE}$ (*44*). During the development of KHI, the upstream edge of the vortex becomes progressively thinner, ultimately forming a sharp, compressional boundary. The observed compressional upstream edge also suggests that the KH waves have progressed beyond the linear surface wave stage and have at least undergone steepening (*45*). At the same time, the strong twisting and folding of field lines in the upstream edge generate the characteristic bipolar magnetic field signatures (*46*–*49*). Moreover, if the cloud is indeed a KH wave, then the magnetic component perpendicular to the flow vortex plane is expected to be amplified and become the dominant component in the upstream edge region due to compression (*46*). In our case, the is corresponds to $B_y^{MSE}$, which is consistent with observations. In addition, the associated pressure variations are also consistent with the feature of KHI: centrifugal forces within the KH wave drive plasma outward, creating a local minimum in total pressure near the center and a pressure peak at the compressional boundary (*50*).

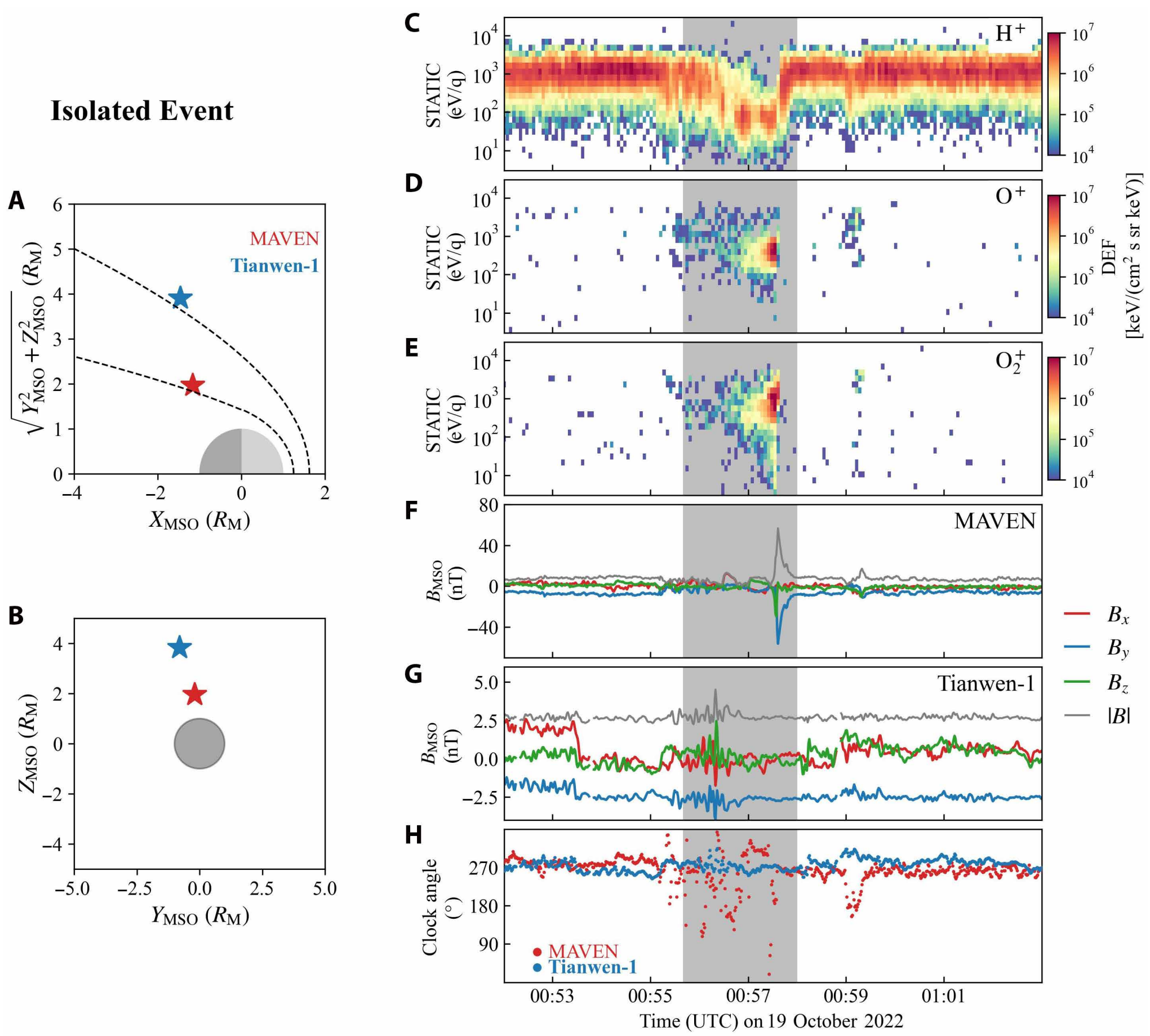

**Fig. 2. Overview of a representative isolated plasma cloud event observed on 19 October 2022.** (**A**) Average spacecraft locations in the $X_{MSO} - \sqrt{Y_{MSO}^2 + Z_{MSO}^2}$ plane. (**B**) Average spacecraft locations in the $YZ_{MSO}$ plane. (**C**) $H^+$ energy spectra. (**D**) $O^+$ energy spectra. (**E**) $O_2^+$ energy spectra. (**F**) Magnetic fields observed by MAVEN in MSO coordinates. (**G**) Magnetic fields observed by Tianwen-1. (**H**) Magnetic field clock angles. The gray shaded interval marks the time interval of the isolated plasma cloud.

We also note a pronounced perturbation in the $V_y^{MSE}$ component during the quasi-periodic event (see Fig. 3D1). This is likely because the event occurred in the flank region, at $Y_{MSE} \sim -1R_M$ (see fig. S1), indicating that the spacecraft intersected the flanks of the KH wave. In contrast, the isolated case was located near $Y_{MSE} \sim 0$, where the spacecraft passed closer to the wave center, and consequently, no clear $V_y^{MSE}$ perturbations were observed. Together, these case studies suggest that, whether quasi-periodic or isolated, plasma clouds exhibit magnetic field and velocity signatures that closely resemble those of nonlinear wave packets driven by the KHI.

## Spatial distribution of plasma clouds and the associated escaping flux

To comprehensively investigate plasma clouds and place them in a global context, we conducted a statistical study using MAVEN and Tianwen-1 datasets from December 2021 to December 2023, when both missions provided overlapping coverage. We first identified intervals during which MAVEN crossed the ICB and selected time windows extending 30 min before and after each crossing. Plasma cloud events were then visually identified within these intervals based on their characteristic signatures in the ion energy spectra and magnetic field observations.

To ensure reliable upstream context, we required that, at the time of each plasma cloud event, Tianwen-1 was located either in the upstream solar wind or in the magnetosheath. When Tianwen-1 was in the solar wind, its magnetic field measurements were directly used to construct the MSE coordinates. When Tianwen-1 was located in the magnetosheath, where local structures can distort the magnetic field, we applied an additional selection criterion: Only events for which the magnetic field clock angles measured by MAVEN and Tianwen-1 differed by less than 30° were retained. This criterion ensures the reliability of the constructed MSE coordinates.

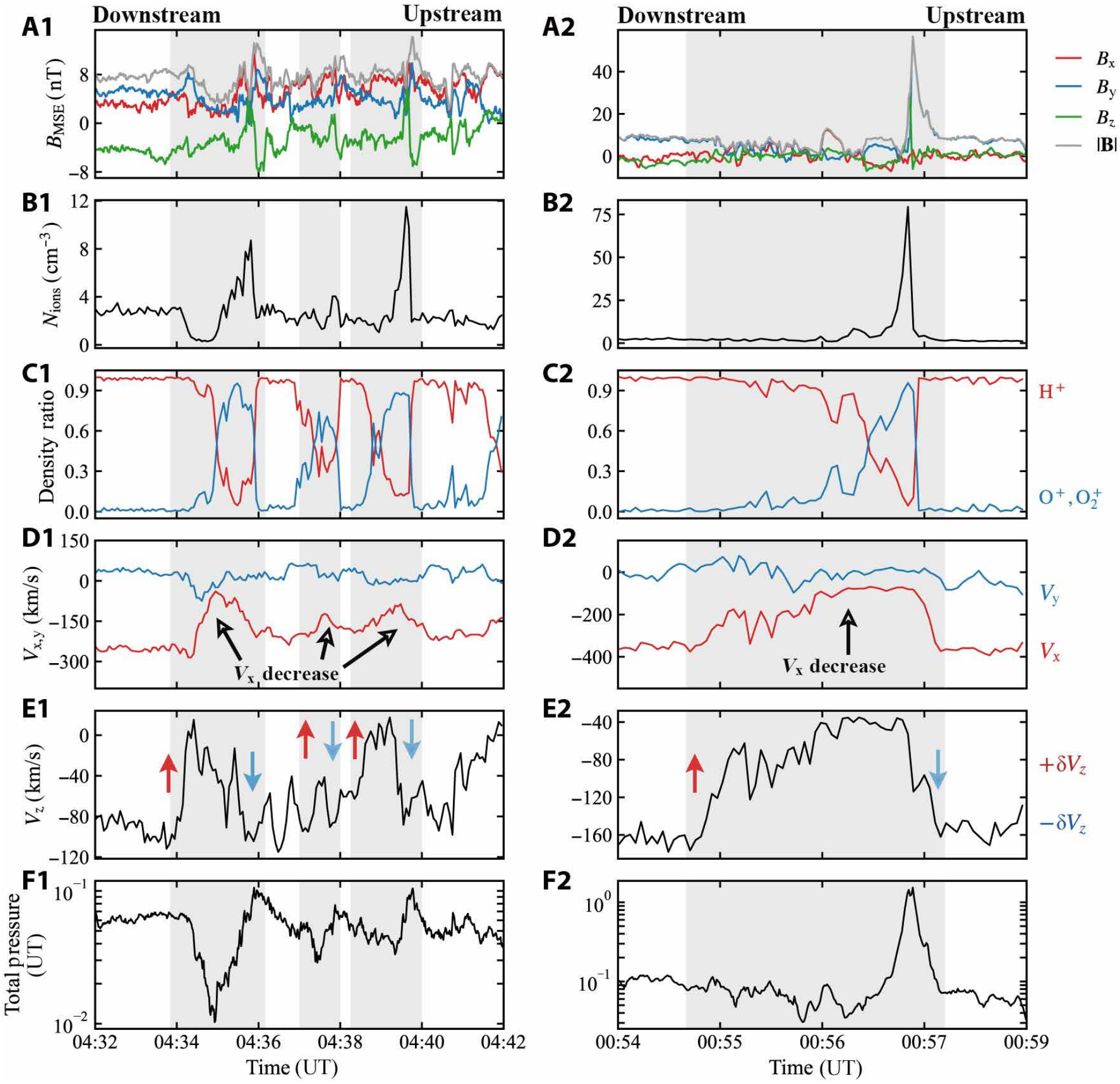

**Fig. 3. Properties of plasma clouds in the MSE coordinates.** (**A1** and **A2**) Magnetic fields. (**B1** and **B2**) Total ion density. (**C1** and **C2**) Density ratio of protons to heavy ions ($O^+$ and $O_2^+$). (**D1** and **D2**) Proton velocities Vx and Vy. (**E1** and **E2**) Proton $V_z$ component ($V_z^{MSE}$). Red (blue) arrows indicate intervals, where $\delta V_z^{MSE} > 0$ ($\delta V_z^{MSE} < 0$), associated with the cloud, where $\delta V_z^{MSE}$ denotes the change in $V_z^{MSE}$. (**F1** and **F2**) total pressure ($P_t$), defined as the sum of plasma thermal pressure and magnetic pressure. The gray shaded regions in panels (A1 to E2) mark the intervals of individual clouds.

In total, we identified 16 isolated events and 46 quasi-periodic events. After transforming into MSE coordinates, all events exhibited magnetic field and plasma signatures consistent with the representative case shown in Figs. 1 to 3. This suggests that these events are also likely driven by the KHI, as illustrated in Fig. 4.

The statistical variations of plasma and magnetic field properties associated with these clouds are summarized in fig. S4. Within the cloud centers (fig. S4, A to C), the total pressure shows a substantial decrease of ~30 to 90%, with a peak around ~50%. The proton $V_x^{MSE}$ also decreases, typically by ~20 to 90%. In contrast, $V_z^{MSE}$ shows much larger variations, peaking around ~100% and reaching up to ~200%, indicating that $V_z^{MSE}$ can decrease to near zero or even reverse sign, consistent with the behavior observed in Fig. 3 and the schematic shown in Fig. 4.

At the upstream edge of the clouds (fig. S4, D to F), we observe the compressional signatures. The total pressure increases by ~50 to 250%, accompanied by an enhancement in magnetic field strength (fig. S4E). In addition, the magnetic field at the upstream edge is dominated by the $B_y^{MSE}$ component (typically above 0.5; see fig. S4F). This is consistent with an enhanced field component perpendicular to the KH plane ($XZ_{MSE}$ plane).

Figure 5 (A and B) shows the spatial distributions of the identified events in the $XR_{MSE}$, and $YZ_{MSE}$ planes, respectively, where $R_{MSE} = \sqrt{Y_{MSE}^2 + Z_{MSE}^2}$. The events are concentrated near the terminator region, corresponding to solar zenith angles between 45° and 135°. Figure 5B further shows that all events occurred in the −E hemisphere and were primarily confined within $|Y_{MSE}| < 1R_M$ region, indicating that they are concentrated near the draping center of the magnetic field lines, where the induced current sheet is located. This indicates that the formation of plasma clouds is strongly influenced by the upstream IMF orientation and the direction of $\vec{E}_{SW}$. Figure 5C demonstrates that these events occur in both regions of strong crustal fields and regions with weak or negligible crustal fields, implying that crustal fields do not play a controlling role in their formation.

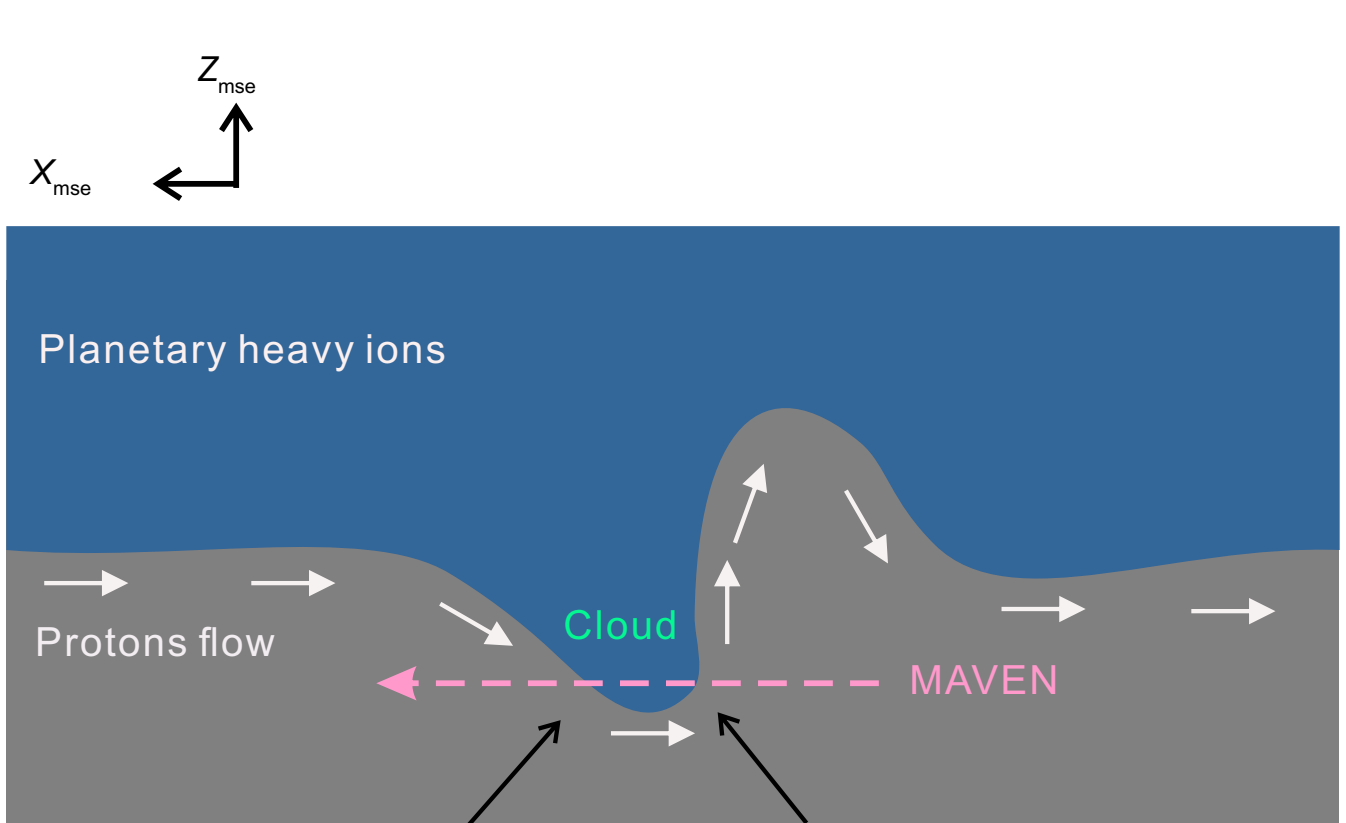

**Fig. 4. Simple schematic illustration of the KHI at the interface between magnetosheath protons and planetary heavy ions.** Protons with higher tailward speeds occupy the lower side, while planetary heavy ions with lower speeds are located on the upper side. Black arrows indicate proton flows near the interface. The pink dashed line with an arrow shows the relative trajectory of MAVEN, which detected a KHI-driven bulge-like expansion of heavy ions, hereafter referred to as a plasma cloud. The first boundary of the cloud encountered by MAVEN is defined as the downstream edge, whereas the boundary where MAVEN exits the cloud is defined as the upstream edge.

Figure 5D presents histograms of the average escaping flux of $O^+$ and $O_2^+$ for these clouds. For each event, we selected data points where the heavy ion density ratio exceeds 0.5 and calculated the mean escaping flux across those points, yielding one representative value per cloud event. We also examined the FOV for each event and excluded those in which the majority of ions were outside the FOV, as their fluxes would be underestimated. As shown in Fig. 5D, we see that most of the average escaping fluxes fall within the range of $10^6$ to $10^8$ $cm^{-2}$ $s^{-1}$, with mean values of $5.04 \times 10^7$ $cm^{-2}$ $s^{-1}$ for $O^+$ and $3.62 \times 10^7$ $cm^{-2}$ $s^{-1}$ for $O_2^+$. We also examined the maximum escaping flux for each event, which yields peak values around ~$10^8$ $cm^{-2}$ $s^{-1}$ for both $O^+$ and $O_2^+$. These results indicate that the escaping flux carried by the plasma clouds is generally on the order of $10^7$ to $10^8$ $cm^{-2}$ $s^{-1}$. For comparison, the average ion escape flux in the magnetotail is ~$1.2 \times 10^6$ $cm^{-2}$ $s^{-1}$ (*21*, *51*), while that in the dayside plume region is about $3.6 \times 10^5$ $cm^{-2}$ $s^{-1}$ (*18*). Therefore, the ion fluxes associated with plasma clouds are roughly one to two orders of magnitude higher than those in the plume region and in the magnetotail.

## DISCUSSION

### Origin of the distinct spatial distribution of plasma clouds

An important question is why the observed plasma clouds are concentrated in the −E hemisphere and predominantly confined within draping center ($|Y_{MSE}| < 1R_M$)? Previous single-spacecraft observations also reported that plasma clouds occur preferentially in the −E hemisphere (*27*), and subsequent hybrid simulations further demonstrated that KHI tends to develop on this side due to diamagnetic drift motion (*30*). According to ideal single-fluid magnetohydrodynamics (MHD) theory (*52*), the growth rate of KHI is

$$\gamma^2 = \frac{\rho_1\rho_2}{(\rho_1+\rho_2)^2}\left(\Delta\vec{V}\cdot\vec{k}\right)^2 - \frac{1}{\mu_0(\rho_1+\rho_2)}\left[\left(\vec{k}\cdot\vec{B}_1\right)^2+\left(\vec{k}\cdot\vec{B}_2\right)^2\right] \quad (1)$$

where $\gamma$ is the KHI growth rate, $\Delta\vec{V}$ is the velocity shear, $\rho_1, \rho_2$ are the mass densities in two regions, $\vec{k}$ is the wave vector of the KHI, $\vec{B}_1, \vec{B}_2$ are the magnetic fields in the two regions, $\mu_0$ is the permeability. Equation 1 indicates that KHI growth is favored by stronger velocity shear ($\Delta\vec{V}$), weaker magnetic fields ($\vec{B}_1$, $\vec{B}_2$), and smaller field-wave alignment ($\vec{B}\cdot\vec{k}$).

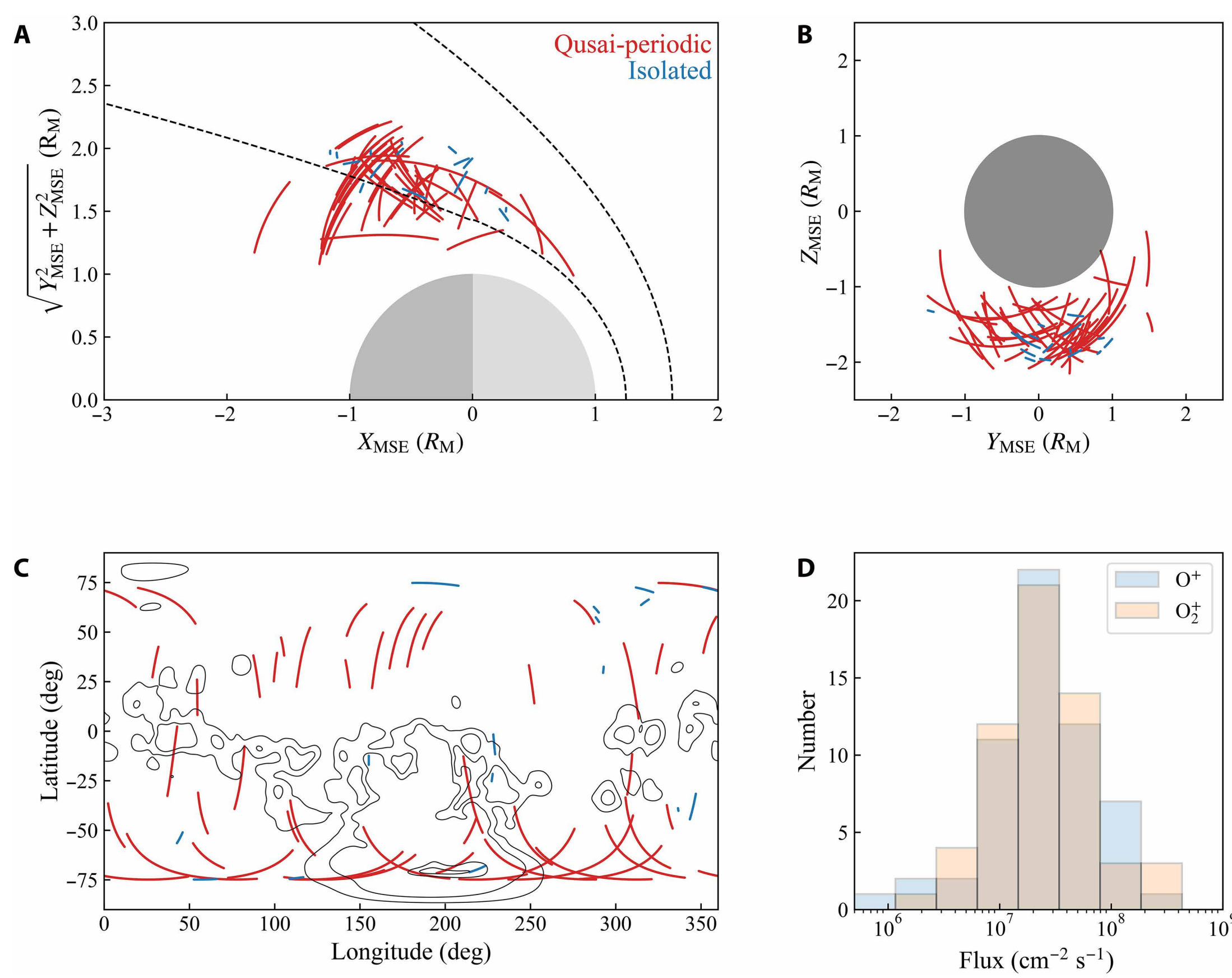

**Fig. 5. Spatial distribution of plasma clouds and the associated escaping ion fluxes.** (**A**) and (**B**) show the distribution of identified cloud events in the $XR_{MSE}$ plane, and $YZ_{MSE}$ plane, respectively. Black curves in (A) indicate the nominal positions of the bow shock and ICB (*41*). The gray sphere in (B) denotes the planetary shadow. (**C**) Distribution of cloud events in planetographic coordinates; black contour lines represent crustal fields from the model of Gao *et al.* (*77*). (**D**) Histogram of $O^+$ and $O_2^+$ fluxes within the clouds.

Zhang *et al.* (*34*) demonstrated that the average spatial distributions of ion density, ion velocity, and magnetic field in MSE coordinates exhibit a pronounced E-hemispheric asymmetry. This suggests that the observed asymmetry in KHI development may be closely tied to the asymmetric distribution of both plasma and magnetic field conditions. Building on these results, we calculate the center-of-mass velocity of heavy ions as

$$\vec{V}_{O,O_2} = \frac{n_O m_o \vec{V}_O + n_{O_2} m_{o_2} \vec{V}_{O_2}}{n_O m_o + n_{O_2} m_{o_2}} \quad (2)$$

where $n$, $m$, and $\vec{V}$ denotes the number density, mass, and velocity. Figure 6A shows that $V_{O,O_2}$ points upward and tailward in both the +E and −E hemispheres, but with a larger magnitude in the +E hemisphere. This hemispheric difference in $V_{O,O_2}$ is also evident in fig. S5C. The asymmetry is attributed to the acceleration of planetary ions by $\vec{E}_{SW}$. Through momentum exchange, solar wind protons are deflected toward the −E direction, resulting in higher speeds in the −E hemisphere (Fig. 6B and fig. S5F), consistent with Dubinin *et al.* (*53*, *54*). As a consequence, the velocity shear, defined as $\Delta\vec{V} = \vec{V}_H - \vec{V}_{O,O_2}$, is stronger in the −E hemisphere (Fig. 6C and fig. S5I), resulting in a larger growth rate $\gamma$ in −E hemisphere.

From the second term in Eq. 1, weaker magnetic fields and orientations more perpendicular to $\vec{k}$ enhance the KHI growth rate. As shown in Fig. 6D and fig. S5L, the magnetic field strength is stronger in the +E hemisphere than in the −E hemisphere due to mass loading (*55*). Furthermore, Fig. 6E demonstrates that within $|Y_{MSE}| < 1R_M$, the magnetic field is nearly perpendicular to the $\vec{X}_{MSE}$ direction (essentially the direction of $\vec{k}$), implying that $\vec{B}\cdot\vec{k}\sim 0$. Together with the distributions of velocity shear and magnetic fields, these results indicate that KHI has large $\gamma$ in the −E hemisphere and near the draping center, supporting the conclusion that plasma clouds are primarily formed by KHI.

It should be noted that Eq. 1 is derived from a single-fluid MHD framework, whereas KHI at Mars involves a mixture of protons and planetary heavy ions; therefore, a multifluid or kinetic description would be more appropriate. However, there is currently no simple analytical expression for these complex conditions. A similar challenge also exists at Venus. Recent multifluid MHD simulations at Venus show similar asymmetric KHI behavior (*56*), consistent with our results, and further suggest that Eq. 1 can capture the essential physics of the asymmetry. Therefore, we also suggest that Eq. 1 provides a reliable first-order explanation for the observed asymmetry at Mars.

Based on the results presented above, we propose a comprehensive scenario for the formation of the observed plasma clouds and their relationship to the KHI (see fig. S6). Near the ICB region, $\vec{E}_{SW}$ accelerates planetary heavy ions, driving them to high velocities in the +E hemisphere and producing a plume-like structure. Because of momentum exchange, solar wind protons are consequently deflected toward the −E direction and gain higher speeds in the −E hemisphere (*53*). This leads to a stronger velocity shear between the planetary heavy ions and solar wind protons in the −E hemisphere. In addition, the acceleration of heavy ions in the +E hemisphere enhances the magnetic field strength via the mass-loading effect (*57*). As a result, the −E hemisphere exhibits both stronger velocity shear between solar wind protons and planetary heavy ions and weaker magnetic field strength, which is favorable for the development of the KHI. Consequently, the KHI develops asymmetrically, with a clear preference for the −E hemisphere. The resulting KH waves are characterized by compressional leading edges, proton flow vortices, and magnetic field rotations. The KH waves transport large amounts of planetary ions, which manifest observationally as plasma clouds. The E-hemisphere asymmetry in KHI occurrence is also consistent with simulations at Mars (*30*) and Venus (*56*, *58*, *59*), suggesting a common tendency for KHI to develop preferentially in induced magnetospheres.

It should be noted that planetary heavy ions have large gyroradii, which may give rise to strong finite Larmor radius (FLR) effects that can influence the development of the KHI. For the quasi-periodic case shown in Fig. 1, if we assume that the phase velocity of the KHI corresponds to the average mass center velocity of magnetosheath protons and planetary heavy ions, then it is about 125 km/s. Given the observed duration of ~70 s, the estimated wavelength is ~8750 km. The average energies of $O^+$ and $O_2^+$ are ~200 eV, corresponding to gyroradii of ~820 and ~1160 km, respectively, suggesting that FLR effects in this case are likely negligible. However, in the +E hemisphere, where heavy ions are more energetic and have larger gyroradii, FLR effects are expected to affect the KHI development. Previous simulation studies (*60*, *61*) have shown that FLR effects enhance the KH growth rate when $\vec{B}\cdot\vec{\Omega} < 0$ and suppress it when $\vec{B}\cdot\vec{\Omega} > 0$, where $\vec{B}$ is the background magnetic field and $\vec{\Omega}$ represents the flow vorticity. In the MSE frame, $\vec{B}$ is basically along the $+\vec{Y}_{MSE}$ direction in the draping center. The flow vorticity, $\vec{\Omega}$, points along $-\vec{Y}_{MSE}$ in the +E hemisphere and along $+\vec{Y}_{MSE}$ in the −E hemisphere. Consequently, $\vec{B}\cdot\vec{\Omega} > 0$ in the −E hemisphere, and $\vec{B}\cdot\vec{\Omega} < 0$ in the +E hemisphere. Thus, the FLR effect is expected to enhance the KHI growth rate in the −E hemisphere while suppressing it in the +E hemisphere, consistent with our observations.

Previous simulations indicate that KHI growth also depends on the thickness of the velocity shear layer: Thicker layers yield larger growth rates (*62*). They further show that the layer thickens when the velocity gradient direction is parallel to the motional electric field and thins when it is antiparallel. At Mars, this geometry implies that the ICB tends to be thicker in the +E hemisphere and thinner in the −E hemisphere. However, our results show KHI occurring in the −E hemisphere, which is opposite to the expectation. This discrepancy indicates that other factors—such as the magnitude of the velocity shear, magnetic field stabilization, and FLR effects—play a greater role in KHI development than shear layer thickness.

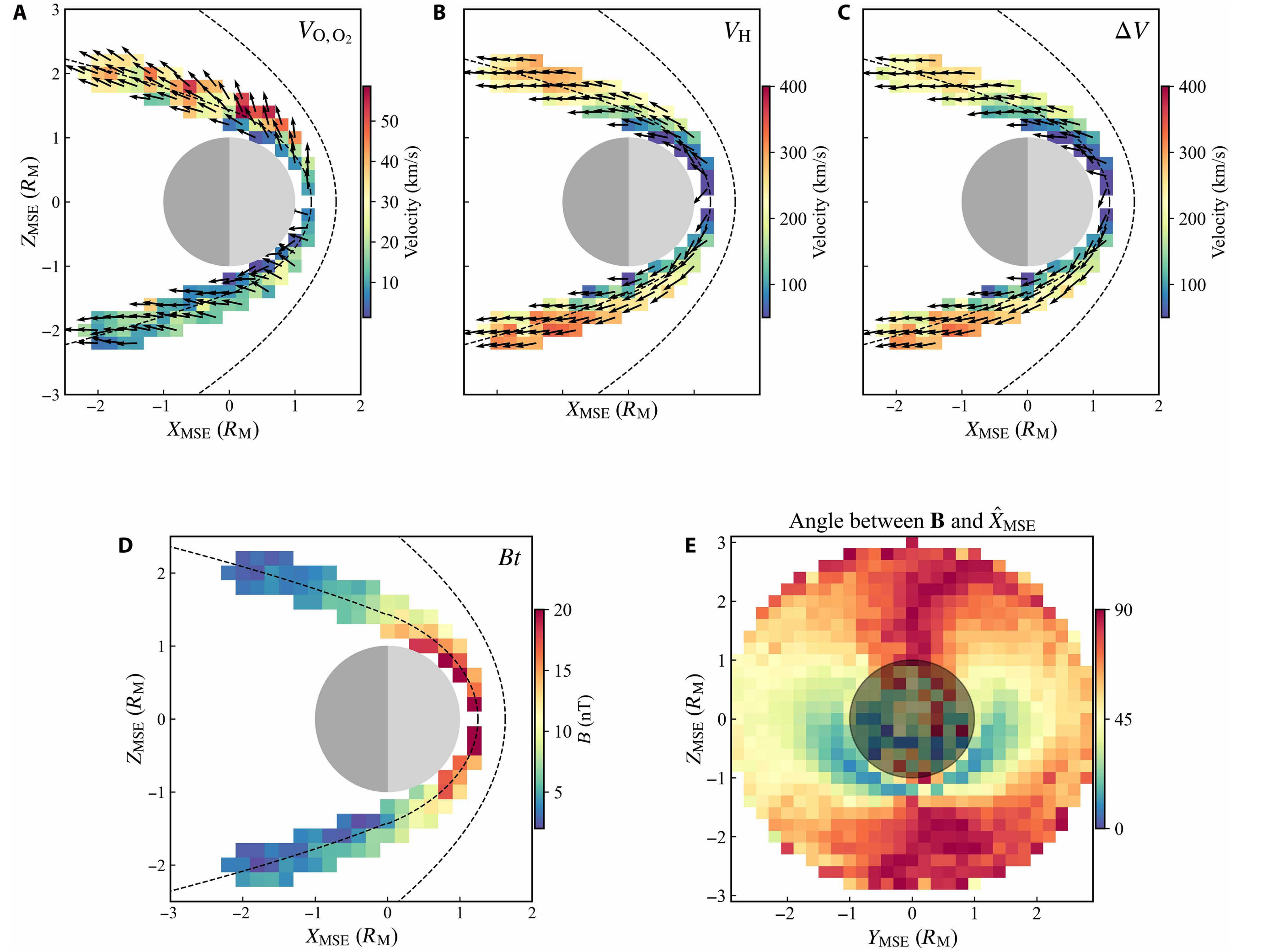

**Fig. 6. Average ion velocity and magnetic field distributions near the ICB.** (**A**) The center-of-mass velocity of heavy ions in the $XZ_{MSE}$ plane, averaged over $-1\ R_M < Y_{MSE} < 1\ R_M$. (**B**) Proton velocity. (**C**) Velocity shear between protons and heavy ions. (**D**) Magnetic field strength. (**E**) Angle between the magnetic field and the $\overline{X}_{MSE}$ in the $YZ_{MSE}$ plane, averaged over $-1.5\ R_M < X_{MSE} < 0.5\ R_M$. Black arrows in (A) to (C) indicate velocity directions. Black curves in (A) to (D) mark the nominal locations of the bow shock and ICB. The black shaded sphere in (E) denotes the planetary shadow.

For intrinsic magnetospheres (Earth, Mercury, and Saturn), KHI also shows hemispheric preferences (*63–65*), yet events are observed in both hemispheres, with different occurrence rates. In contrast, in Mars's induced magnetosphere, we have not identified any events in the +E hemisphere in our dataset, indicating a much stronger hemispheric asymmetry. This suggests that KHI may behave differently in induced magnetospheres than in intrinsic ones. A key difference is the magnetic configuration: The induced magnetic fields are nearly parallel to the adjacent magnetosheath magnetic fields, resembling the configuration of Earth's magnetosphere under northward IMF conditions, where magnetic reconnection is generally suppressed and KHI is more likely to develop. Topology also differs: Induced fields are not anchored to the planet, whereas intrinsic fields are rooted in the planetary polar region. Anchored field lines are more easily distorted and twisted, and they can also exert strong magnetic tension, which can act to suppress KHI growth. In a word, Mars provides a natural laboratory for studying KHI in an induced magnetosphere, including prominent kinetic and multispecies ion effects.

## Origin and nature of isolated plasma clouds

In the scenario described in fig. S6, plasma clouds are generally expected to appear as quasi-periodic structures rather than isolated ones. However, the isolated clouds exhibit observational signatures remarkably similar to those of the quasi-periodic cases, raising an important question regarding the nature of these isolated clouds. One possible explanation is that these structures are detached KH vortices (see fig. S7A). This scenario requires KHI to evolve into its nonlinear stage and form vortices. Within these vortices, the magnetic field lines become strongly distorted, enabling secondary magnetic reconnection to occur (*66–69*), which can ultimately lead to the detachment of the vortices. The observed compressional upstream edge suggests that the KH waves have progressed beyond the linear surface wave stage and have at least undergone steepening. However, whether these waves have fully evolved into rolled-up vortices remains uncertain (*70*), and signatures of magnetic reconnection within the KH structures have yet to be observed.

Another possible explanation is that the so-called isolated clouds may, in fact, be part of a train of KH waves, with the spacecraft sampling only the largest wave crest due to its relatively distant position from the ICB (see fig. S7B). In these cases, surrounding wave structures with smaller amplitudes may go undetected. As seen in our case study, we can also observe weaker cloud-like signatures adjacent to the main event (e.g., at 00:59 in Fig. 1, A2 to E2), supporting the idea that the observed isolated cloud is actually embedded within a broader wave train, while the weaker signatures correspond to lower-amplitude structures within the same train.

Regardless of whether the plasma cloud is quasi-periodic or isolated, we observe intense wave activity within the cloud (see fig. S8), particularly near the upstream compressional boundary. These waves can interact with particles, and these wave-particle interactions likely facilitate the mixing of solar wind and planetary ions, thereby enhancing mass and energy transport and exchange between the two populations.

### Role in ion escape

Previous studies have reported ion escape rates associated with KHI or plasma clouds spanning several orders of magnitude, from $2 \times 10^{23}$ to $3 \times 10^{24}$ s$^{-1}$ (*25*, *26*) to $1.5 \times 10^{24}$ s$^{-1}$ (*30*), $5 \times 10^{25}$ s$^{-1}$ (*31*), and as high as $6 \times 10^{26}$ s$^{-1}$ (*29*). These discrepancies mainly reflect different assumptions about the spatial extent of plasma clouds. Simultaneous two-point measurements provide a way to directly constrain their spatial scale. Here, we show an example, when MAVEN and Tianwen-1 were separated by only 0.56 $R_M$, MAVEN observed an isolated cloud but Tianwen-1 did not (fig. S9). This suggests that the spatial scale of plasma clouds can vary widely and can be much smaller than previous estimates of 2.5 to 6 $R_M$ (*29*–*31*).

Determining their lifetime is more challenging, as spacecraft cannot remain the ICB for long periods. Previous escape-rate estimates often assumed that plasma clouds occur continuously, but our results show that their occurrence is strongly controlled by the orientation of $\vec{E}_{SW}$. Detection therefore requires a favorable spacecraft position. Nevertheless, the fact that both the preceding and following MAVEN orbits of the quasi-periodic event analyzed in Results showed plasma cloud signatures near the ICB (fig. S10) suggests that plasma clouds may be generated continuously or that the KHI may persist for durations of at least 8 hours. Collectively, these findings imply that planetary ion escape could be continuously modulated by the KHI. Under a simple geometric assumption, where a nonlinear KH wave packet is treated as a cylinder with a radius of $\sim 0.5\, R_M$, the observed total flux of $O_2^+$ and $O^+$ ($\sim 10^8$ cm$^{-2}$ s$^{-1}$) implies a total escape rate of $\sim 10^{25}$ s$^{-1}$. This rate is an order of magnitude higher than that of the plume escape channel and is comparable to the tail escape channel.

Consistent with the spatial distribution of KHI, previous statistical studies [e.g., (*19*)] have shown that heavy ion escape fluxes in the Martian magnetotail are substantially enhanced in the −E hemisphere. However, this asymmetry is unlikely to be caused by KHI. Instead, it is fundamentally controlled by the decoupling between solar wind protons and planetary ions [e.g., (*71*)]. Consistently, multifluid MHD simulations, even without resolving KHI, reproduce the same −E-preferred escape pattern (*15*).

Nevertheless, this does not diminish the importance of KHI in ion escape processes. In the event analyzed here, $O_2^+$ reach energies up to ~1 keV, comparable to plume ions (*17*), but exhibit higher temperatures within KH wave packets than typical plume ions. This indicates that KHI plays an important role in both heating and acceleration of ionospheric ions. Moreover, KHI can strongly modulate the spatial structure of ion escape in the −E hemisphere. In particular, KHI-induced structures can organize escaping ions into wave-like or detached plasma clouds, producing banded or intermittent features in the escaping ion fluxes (*30*).

In summary, we investigate a distinct mode of bulk ion escape at Mars, known as plasma clouds, using joint observations from MAVEN and Tianwen-1. The availability of real-time IMF measurements enables accurate characterization of the background induced magnetic fields and plasma flows. This, in turn, allows for detailed analysis of plasma cloud properties and offers new insights into their underlying formation mechanisms. The key findings can be summarized as follows

1) These plasma clouds feature vortex-like proton flows and a central depletion in both magnetic field and total pressure, followed by a sharp enhancement at the upstream edge. This observed structure is consistent with nonlinear wave packets originating from the KHI.

2) The plasma clouds are primarily distributed near the terminator and the nightside ICB, particularly around the draping center, and occur exclusively in the −E hemisphere. This spatial pattern is attributed to the enhanced KHI growth rates in these regions, where the background induced magnetic fields and plasma flow shear are most favorable for instability. This also suggests that plasma clouds arise naturally from the solar wind interaction with Mars's upper atmosphere, and similar features may be common at other unmagnetized planets such as Venus.

3) Plasma clouds exhibit $O^+$ and $O_2^+$ escape fluxes in the range of $10^6$ to $10^8$ cm$^{-2}$ s$^{-1}$, with mean values of $5.04 \times 10^7$ cm$^{-2}$ s$^{-1}$ for $O^+$ and $3.62 \times 10^7$ cm$^{-2}$ s$^{-1}$ for $O_2^+$, about one to two orders of magnitude higher than typical steady escaping channels (plume and magnetotail). This indicates that plasma clouds, or the KHI, represent an important pathway for atmospheric ion loss.

4) Leveraging the advantages of joint observations, we are able to constrain the spatial scale of plasma clouds or KH wave packets for the first time. Our results show that their spatial extent can be smaller than 0.6 $R_M$ (~2000 km), much smaller than previous estimates based on single-point measurements. This highlights the importance of multipoint observations in advancing our understanding of magnetospheric processes and the associated atmospheric ion escape.

Based on these findings, we show that KHI facilitates mass and energy transport not only within intrinsic magnetospheres but also within induced magnetospheres such as those of Mars and Venus, where it drives exchange between the solar wind and planetary upper atmospheres. This underscores the fundamental role of KHI in shaping solar wind–planet interactions across both magnetized and unmagnetized environments.

## MATERIALS AND METHODS

We adopt ion measurements from the Suprathermal and Thermal Ion Composition [STATIC; (*72*)] instrument, and magnetic field measurements from the Magnetometer [MAG; (*73*)]. The MAG is a fluxgate magnetometer that measures three-dimensional magnetic field vectors at a frequency of 32 Hz. We also use the c6 and d1 data of STATIC that provides the omni spectrum and three-dimensional velocity distributions of $H^+$, $O^+$, and $O_2^+$ with energy between 0 and 30 keV/q at time resolution of 4 s. STATIC consists of a time-of-flight sensor that can measure the mass-per-charge of ions and then determine the ion species. The FOV of STATIC is 360° × 90°. In addition to MAVEN, we use the magnetic field data measured by the Tianwen-1 Mars Orbiter Magnetometer [MOMAG; (*74*–*76*)]. The MOMAG could measure the three-dimensional magnetic field vectors at a frequency of 128 Hz. For our analysis, we use data products that are provided at a time resolution of 1 Hz.

### MSO and MSE coordinates

We employed two Cartesian coordinate systems. The first is the MSO coordinates. Here, $\vec{X}_{MSO}$ is along the vector from Mars to the Sun, $\vec{Z}_{MSO}$ is perpendicular to the orbital plane, and $\vec{Y}_{MSO}$ completes the right-handed system, closely aligned with the opposite direction of the orbital velocity vector. To better resolve the magnetic and velocity signatures of the plasma cloud, and given that the configuration of the induced magnetosphere is strongly influenced by the solar wind and IMF orientation, we also adopted the MSE coordinate system. In this system, $\vec{X}_{MSE}$ is antiparallel to the upstream solar wind velocity, $\vec{Z}_{MSE}$ aligns with the direction of the upstream solar wind convective electric field ($\vec{E}_{SW}$), and $\vec{Y}_{MSE}$ completes the right-handed system, aligning with the cross-flow magnetic field component of the IMF. Thus, the $Z_{MSE} > 0$ hemisphere corresponds to the +E hemisphere, while $Z_{MSE} < 0$ denotes the −E hemisphere. Throughout this paper, we assume that the upstream solar wind flow is purely along the $-\vec{X}_{MSO}$ direction, which implies that $\vec{X}_{MSO}$ and $\vec{X}_{MSE}$ are equivalent.

## Supplementary Materials

**This PDF file includes:**

Supplementary Text

Figs. S1 to S10

## REFERENCES

1. J. A. Hurowitz, J. P. Grotzinger, W. W. Fischer, S. M. McLennan, R. E. Milliken, N. Stein, A. R. Vasavada, D. F. Blake, E. Dehouck, J. L. Eigenbrode, A. G. Fairén, J. Frydenvang, R. Gellert, J. A. Grant, S. Gupta, K. E. Herkenhoff, D. W. Ming, E. B. Rampe, M. E. Schmidt, K. L. Siebach, K. Stack-Morgan, D. Y. Sumner, R. C. Wiens, Redox stratification of an ancient lake in Gale crater, Mars. *Science* **356**, eaah6849 (2017).
2. B. M. Hynek, M. Beach, M. R. T. Hoke, Updated global map of Martian valley networks and implications for climate and hydrologic processes. *J. Geophys. Res.* **115**, E09008 (2010).
3. R. J. Lillis, S. Robbins, M. Manga, J. S. Halekas, H. V. Frey, Time history of the Martian dynamo from crater magnetic field analysis. *J. Geophys. Res. Planets* **118**, 1488–1511 (2013).
4. B. Jakosky, D. Brain, M. Chaffin, S. Curry, J. Deighan, J. Grebowsky, J. Halekas, F. Leblanc, R. Lillis, J. Luhmann, L. Andersson, N. Andre, D. Andrews, D. Baird, D. Baker, J. Bell, M. Benna, D. Bhattacharyya, S. Bougher, C. Bowers, P. Chamberlin, J. Y. Chaufray, J. Clarke, G. Collinson, M. Combi, J. Connerney, K. Connour, J. Correira, K. Crabb, F. Crary, T. Cravens, M. Crismani, G. Delory, R. Dewey, G. DiBraccio, C. Dong, Y. Dong, P. Dunn, H. Egan, M. Elrod, S. England, F. Eparvier, R. Ergun, A. Eriksson, T. Esman, J. Espley, S. Evans, K. Fallows, X. Fang, M. Fillingim, C. Flynn, A. Fogle, C. Fowler, J. Fox, M. Fujimoto, P. Garnier, Z. Girazian, H. Groeller, J. Gruesbeck, O. Hamil, K. Hanley, T. Hara, Y. Harada, J. Hermann, M. Holmberg, G. Holsclaw, S. Houston, S. Inui, S. Jain, R. Jolitz, A. Kotova, T. Kuroda, D. Larson, Y. Lee, C. Lee, F. Lefevre, C. Lentz, D. Lo, R. Lugo, Y. J. Ma, P. Mahaffy, M. Marquette, Y. Matsumoto, M. Mayyasi, C. Mazelle, W. McClintock, J. McFadden, A. Medvedev, M. Mendillo, K. Meziane, Z. Milby, D. Mitchell, R. Modolo, F. Montmessin, A. Nagy, H. Nakagawa, C. Narvaez, K. Olsen, D. Pawlowski, W. Peterson, A. Rahmati, K. Roeten, N. Romanelli, S. Ruhunusiri, C. Russell, S. Sakai, N. Schneider, K. Seki, R. Sharrar, S. Shaver, D. Siskind, M. Slipski, Y. Soobiah, M. Steckiewicz, M. Stevens, I. Stewart, A. Stiepen, S. Stone, V. Tenishev, N. Terada, K. Terada, E. Thiemann, R. Tolson, G. Toth, J. Trovato, M. Vogt, T. Weber, P. Withers, S. Xu, R. Yelle, E. Yiğit, R. Zurek, Loss of the Martian atmosphere to space: Present-day loss rates determined from MAVEN observations and integrated loss through time. *Icarus* **315**, 146–157 (2018).
5. C. Dong, Y. Lee, Y. Ma, M. Lingam, S. Bougher, J. Luhmann, S. Curry, G. Toth, A. Nagy, V. Tenishev, X. Fang, D. Mitchell, D. Brain, B. Jakosky, Modeling martian atmospheric losses over Time: Implications for exoplanetary climate evolution and habitability. *Astrophys. J. Lett.* **859**, L14 (2018).
6. R. J. Lillis, D. A. Brain, S. W. Bougher, F. Leblanc, J. G. Luhmann, B. M. Jakosky, R. Modolo, J. Fox, J. Deighan, X. Fang, Y. C. Wang, Y. Lee, C. Dong, Y. Ma, T. Cravens, L. Andersson, S. M. Curry, N. Schneider, M. Combi, I. Stewart, J. Clarke, J. Grebowsky, D. L. Mitchell, R. Yelle, A. F. Nagy, D. Baker, R. P. Lin, Characterizing atmospheric escape from Mars today and through time, with MAVEN. *Space Sci. Rev.* **195**, 357–422 (2015).
7. J. L. Fox, A. B. Hać, Photochemical escape of oxygen from Mars: A comparison of the exobase approximation to a Monte Carlo method. *Icarus* **204**, 527–544 (2009).
8. T. E. Cravens, A. Rahmati, J. L. Fox, R. Lillis, S. Bougher, J. Luhmann, S. Sakai, J. Deighan, Y. Lee, M. Combi, B. Jakosky, Hot oxygen escape from Mars: Simple scaling with solar EUV irradiance. *J. Geophys. Res. Space Phys.* **122**, 1102–1116 (2017).
9. R. J. Lillis, J. Deighan, J. L. Fox, S. W. Bougher, Y. Lee, M. R. Combi, T. E. Cravens, A. Rahmati, P. R. Mahaffy, M. Benna, M. K. Elrod, J. P. McFadden, R. E. Ergun, L. Andersson, C. M. Fowler, B. M. Jakosky, E. Thiemann, F. Eparvier, J. S. Halekas, F. Leblanc, J. Chaufray, Photochemical escape of oxygen from Mars: First results from MAVEN in situ data. *J. Geophys. Res. Space Phys.* **122**, 3815–3836 (2017).
10. T. E. Cravens, O. Hamil, S. Houston, S. Bougher, Y. Ma, D. Brain, S. Ledvina, Estimates of ionospheric transport and ion loss at Mars. *J. Geophys. Res. Space Phys.* **122**, 10626–10637 (2017).
11. R. Lundin, A. Zakharov, R. Pellinen, H. Borg, B. Hultqvist, N. Pissarenko, E. M. Dubinin, S. W. Barabash, I. Liede, H. Koskinen, First measurements of the ionospheric plasma escape from Mars. *Nature* **341**, 609–612 (1989).
12. C. Zhang, Z. Rong, X. Li, M. Fränz, H. Nilsson, R. Jarvinen, M. Persson, Y. Futaana, C. Dong, M. Yamauchi, J. Gao, Y. Zhou, L. Wang, Z. Shi, Y. Wei, F. He, M. Holmström, S. Barabash, The energetic oxygen ion beams in the martian magnetotail current sheets: Hints from the comparisons between two types of current sheets. *Geophys. Res. Lett.* **51**, e2023GL107190 (2024).
13. C. Zhang, C. Dong, H. Zhou, J. Halekas, M. Yamauchi, H. Nilsson, T. Z. Liu, M. Persson, S. Curry, Y. Dong, Y. Futaana, Y. Chen, M. Zhou, R. Suranga, K. G. Hanley, C. Mazelle, S. Xu, R. Ramstad, M. Holmström, L. J. Chen, Anomalous transient enhancement of planetary ion escape at Mars. *Nat. Commun.* **16**, 3159 (2025).
14. C. Dong, S. W. Bougher, Y. Ma, G. Toth, Y. Lee, A. F. Nagy, V. Tenishev, D. J. Pawlowski, M. R. Combi, D. Najib, Solar wind interaction with the Martian upper atmosphere: Crustal field orientation, solar cycle, and seasonal variations. *J. Geophys. Res. Space Phys.* **120**, 7857–7872 (2015).
15. C. Dong, S. W. Bougher, Y. Ma, G. Toth, A. F. Nagy, D. Najib, Solar wind interaction with Mars upper atmosphere: Results from the one-way coupling between the multifluid MHD model and the MTGCM model. *Geophys. Res. Lett.* **41**, 2708–2715 (2014).
16. H. Nilsson, Q. Zhang, G. Stenberg Wieser, M. Holmström, S. Barabash, Y. Futaana, A. Fedorov, M. Persson, M. Wieser, Solar cycle variation of ion escape from Mars. *Icarus* **393**, 114610 (2023).
17. R. Ramstad, S. Barabash, Y. Futaana, H. Nilsson, M. Holmström, Ion escape from mars through time: An extrapolation of atmospheric loss based on 10 years of mars express measurements. *J. Geophys. Res. Planets* **123**, 3051–3060 (2018).
18. Y. Dong, X. Fang, D. A. Brain, J. P. McFadden, J. S. Halekas, J. E. Connerney, S. M. Curry, Y. Harada, J. G. Luhmann, B. M. Jakosky, Strong plume fluxes at Mars observed by MAVEN: An important planetary ion escape channel. *Geophys. Res. Lett.* **42**, 8942–8950 (2015).
19. X. Z. Li, Z. J. Rong, M. Fraenz, C. Zhang, L. Klinger, Z. Shi, J. W. Gao, M. W. Dunlop, Y. Wei, Two types of martian magnetotail current sheets: MAVEN observations of ion composition. *Geophys. Res. Lett.* **50**, e2022GL102630 (2023).
20. E. Dubinin, M. Fraenz, M. Pätzold, J. McFadden, J. S. Halekas, G. A. DiBraccio, J. E. P. Connerney, F. Eparvier, D. Brain, B. M. Jakosky, O. Vaisberg, L. Zelenyi, The effect of solar wind variations on the escape of oxygen ions from mars through different channels: MAVEN observations. *J. Geophys. Res. Space Phys.* **122**, 11285–11301 (2017).
21. S. Inui, K. Seki, S. Sakai, D. Brain, T. Hara, J. McFadden, J. Halekas, D. Mitchell, G. DiBraccio, B. Jakosky, Statistical Study of Heavy Ion Outflows From Mars Observed in the Martian-Induced Magnetotail by MAVEN. *J. Geophys. Res. Space Phys.* **124**, 5482–5497 (2019).
22. L. Brace, R. Theis, W. Hoegy, Plasma clouds above the ionopause of Venus and their implications. *Planet. Space Sci.* **30**, 29–37 (1982).
23. M. H. Acuña, J. E. P. Connerney, P. Wasilewski, R. P. Lin, K. A. Anderson, C. W. Carlson, J. McFadden, D. W. Curtis, D. Mitchell, H. Reme, C. Mazelle, J. A. Sauvaud, C. d'Uston, A. Cros, J. L. Medale, S. J. Bauer, P. Cloutier, M. Mayhew, D. Winterhalter, N. F. Ness, Magnetic field and plasma observations at Mars: Initial results of the Mars Global Surveyor mission. *Science* **279**, 1676–1680 (1998).
24. C. T. Russell, J. G. Luhmann, R. C. Elphic, F. L. Scarf, L. H. Brace, Magnetic field and plasma wave observations in a plasma cloud at Venus. *Geophys. Res. Lett.* **9**, 45–48 (1982).

25. T. Penz, N. Erkaev, H. Biernat, H. Lammer, U. Amerstorfer, H. Gunell, E. Kallio, S. Barabash, S. Orsini, A. Milillo, W. Baumjohann, Ion loss on Mars caused by the Kelvin–Helmholtz instability. *Planet. Space Sci.* **52**, 1157–1167 (2004).
26. J. S. Halekas, D. A. Brain, S. Ruhunusiri, J. P. McFadden, D. L. Mitchell, C. Mazelle, J. E. P. Connerney, Y. Harada, T. Hara, J. R. Espley, G. DiBraccio, B. M. Jakosky, Plasma clouds and snowplows: Bulk plasma escape from Mars observed by MAVEN. *Geophys. Res. Lett.* **43**, 1426–1434 (2016).
27. J. S. Halekas, S. Ruhunusiri, J. P. McFadden, J. R. Espley, G. A. DiBraccio, Ion composition boundary layer instabilities at Mars. *Geophys. Res. Lett.* **46**, 10303–10312 (2019).
28. C. Zhang, Z. Rong, H. Nilsson, L. Klinger, S. Xu, Y. Futaana, Y. Wei, J. Zhong, M. Fränz, K. Li, H. Zhang, K. Fan, L. Wang, M. Holmström, Y. Ge, J. Cui, MAVEN Observations of Periodic Low-altitude Plasma Clouds at Mars. *Astrophys. J. Lett.* **922**, L33 (2021).
29. G. Poh, J. R. Espley, K. Nykyri, C. M. Fowler, X. Ma, S. Xu, G. Hanley, N. Romanelli, C. Bowers, J. Gruesbeck, G. A. DiBraccio, On the growth and development of non-linear Kelvin–Helmholtz instability at Mars: MAVEN observations. *J. Geophys. Res. Space Phys.* **126**, e2021JA029224 (2021).
30. L. Wang, C. Huang, A. Du, Y. Ge, G. Chen, Z. Yang, S. Li, K. Zhang, Kelvin–Helmholtz instability at Mars: In situ observations and kinetic simulations. *Astrophys. J.* **947**, 51 (2023).
31. X. Wang, X. Xu, Y. Ye, J. Wang, M. Wang, Z. Zhou, Q. Chang, Q. Xu, J. Xu, L. Luo, P. He, S. Cheng, MAVEN observations of the Kelvin-Helmholtz instability developing at the ionopause of Mars. *Geophys. Res. Lett.* **49**, e2022GL098673 (2022).
32. Z. Koh, G. Poh, C. M. Fowler, K. G. Hanley, X. Ma, J. R. Gruesbeck, D. C. P. Kuruppuaratchi, W. Sun, G. A. DiBraccio, J. R. Espley, Global occurrence of Kelvin-Helmholtz vortices at Mars. *Geophys. Res. Lett.* **52**, e2025GL117836 (2025).
33. C. Zhang, Z. Rong, L. Klinger, H. Nilsson, Z. Shi, F. He, J. Gao, X. Li, Y. Futaana, R. Ramstad, X. Wang, M. Holmström, S. Barabash, K. Fan, Y. Wei, Three-dimensional configuration of induced magnetic fields around Mars. *J. Geophys. Res. Planets* **127**, e2022JE007334 (2022).
34. C. Zhang, C. Dong, H. Zhou, J. Halekas, X. Li, J. Gao, H. Shen, X. Wang, H. Nilsson, R. Ramstad, C. Mazelle, L. Wang, S. Xu, A. Tadlock, K. G. Hanley, S. M. Curry, D. L. Mitchell, Global energy transport and conversion in the solar wind-mars interaction: MAVEN observations. *J. Geophys. Res. Planets* **130**, e2025JE009295 (2025).
35. C. Zhang, C. Dong, H. Zhou, J. Deca, S. Xu, Y. Harada, S. M. Curry, D. L. Mitchell, Z. Liu, J. Qin, C. Mazelle, Observational characteristics of electron distributions in the martian induced magnetotail. *Geophys. Res. Lett.* **52**, e2024GL113030 (2025).
36. J. Luhmann, S. Ledvina, C. Russell, Induced magnetospheres. *Adv. Space Res.* **33**, 1905–1912 (2004).
37. M. L. Marquette, R. J. Lillis, J. S. Halekas, J. G. Luhmann, J. R. Gruesbeck, J. R. Espley, Autocorrelation study of solar wind plasma and IMF Properties as measured by the MAVEN Spacecraft. *J. Geophys. Res. Space Phys.* **123**, 2493–2512 (2018).
38. A. R. Azari, E. Abrahams, F. Sapienza, J. Halekas, J. Biersteker, D. L. Mitchell, F. Pérez, M. Marquette, M. J. Rutala, C. F. Bowers, C. M. Jackman, S. M. Curry, A virtual solar wind monitor at mars with uncertainty quantification using gaussian processes. *J. Geophys. Res. Mach. Learn. Comput.* **1**, e2024JH000155 (2024).
39. W. X. Wan, C. Wang, C. L. Li, Y. Wei, China's first mission to Mars. *Nat. Astron.* **4**, 721–721 (2020).
40. B. M. Jakosky, R. P. Lin, J. M. Grebowsky, J. G. Luhmann, D. F. Mitchell, G. Beutelschies, T. Priser, M. Acuna, L. Andersson, D. Baird, D. Baker, R. Bartlett, M. Benna, S. Bougher, D. Brain, D. Carson, S. Cauffman, P. Chamberlin, J. Y. Chaufray, O. Cheatom, J. Clarke, J. Connerney, T. Cravens, D. Curtis, G. Delory, S. Demcak, A. DeWolfe, F. Eparvier, R. Ergun, A. Eriksson, J. Espley, X. Fang, D. Folta, J. Fox, C. Gomez-Rosa, S. Habenicht, J. Halekas, G. Holsclaw, M. Houghton, R. Howard, M. Jarosz, N. Jedrich, M. Johnson, W. Kasprzak, M. Kelley, T. King, M. Lankton, D. Larson, F. Leblanc, F. Lefevre, R. Lillis, P. Mahaffy, C. Mazelle, W. McClintock, J. McFadden, D. L. Mitchell, F. Montmessin, J. Morrissey, W. Peterson, W. Possel, J. A. Sauvaud, N. Schneider, W. Sidney, S. Sparacino, A. I. F. Stewart, R. Tolson, D. Toublanc, C. Waters, T. Woods, R. Yelle, R. Zurek, The Mars Atmosphere and Volatile Evolution (MAVEN) Mission. *Space Sci. Rev.* **195**, 3–48 (2015).
41. J. Trotignon, C. Mazelle, C. Bertucci, M. Acuña, Martian shock and magnetic pile-up boundary positions and shapes determined from the Phobos 2 and Mars Global Surveyor data sets. *Planet. Space Sci.* **54**, 357–369 (2006).
42. Y. Dong, X. Fang, D. A. Brain, D. M. Hurley, J. S. Halekas, J. R. Espley, R. Ramstad, S. Ruhunusiri, B. M. Jakosky, Magnetic field in the martian magnetosheath and the application as an IMF clock angle proxy. *J. Geophys. Res. Space Phys.* **124**, 4295–4313 (2019).
43. Z. Cheng, C. Zhang, C. Dong, H. Zhou, J. Gao, A. Tadlock, X. Li, L. Wang, Revisiting Mars' induced magnetic field and clock angle departures under real-time upstream solar wind conditions. *J. Geophys. Res. Space Phys.* **130**, e2025JA034688 (2025).
44. A. Masson, K. Nykyri, Kelvin–Helmholtz instability: Lessons learned and ways forward. *Space Sci. Rev.* **214**, 71 (2018).
45. J. R. Johnson, S. Wing, P. A. Delamere, Kelvin Helmholtz Instability in Planetary Magnetospheres. *Space Sci. Rev.* **184**, 1–31 (2014).
46. H. Hasegawa, M. Fujimoto, T. D. Phan, H. Rème, A. Balogh, M. W. Dunlop, C. Hashimoto, R. TanDokoro, Transport of solar wind into Earth's magnetosphere through rolled-up Kelvin–Helmholtz vortices. *Nature* **430**, 755–758 (2004).
47. A. Otto, D. H. Fairfield, Kelvin-Helmholtz instability at the magnetotail boundary: MHD simulation and comparison with Geotail observations. *J. Geophys. Res.* **105**, 21175–21190 (2000).
48. D. H. Fairfield, M. M. Kuznetsova, T. Mukai, T. Nagai, T. I. Gombosi, A. J. Ridley, Waves on the dusk flank boundary layer during very northward interplanetary magnetic field conditions: Observations and simulation. *J. Geophys. Res.* **112**, A08206 (2007).
49. T. Sundberg, S. A. Boardsen, J. A. Slavin, L. G. Blomberg, J. A. Cumnock, S. C. Solomon, B. J. Anderson, H. Korth, Reconstruction of propagating Kelvin–Helmholtz vortices at Mercury's magnetopause. *Planet. Space Sci.* **59**, 2051–2057 (2011).
50. H. Hasegawa, Structure and dynamics of the magnetopause and its boundary layers. *Monogr. Environ. Earth Planets* **1**, 71–119 (2012).
51. D. A. Brain, J. P. McFadden, J. S. Halekas, J. E. P. Connerney, S. W. Bougher, S. Curry, C. F. Dong, Y. Dong, F. Eparvier, X. Fang, K. Fortier, T. Hara, Y. Harada, B. M. Jakosky, R. J. Lillis, R. Livi, J. G. Luhmann, Y. Ma, R. Modolo, K. Seki, The spatial distribution of planetary ion fluxes near Mars observed by MAVEN. *Geophys. Res. Lett.* **42**, 9142–9148 (2015).
52. S. Chandrasekhar, *Hydromagnetic and hydrodynamic stability* (Oxford Univ. Press, 1961).
53. E. Dubinin, M. Fraenz, M. Pätzold, J. S. Halekas, J. Mcfadden, J. E. P. Connerney, B. M. Jakosky, O. Vaisberg, L. Zelenyi, Solar wind deflection by mass loading in the Martian magnetosheath based on MAVEN observations. *Geophys. Res. Lett.* **45**, 2574–2579 (2018).
54. E. Dubinin, M. Fraenz, M. Pätzold, S. Tellmann, J. McFadden, J. Halekas, G. DiBraccio, Solar wind—Ionosphere interface at Mars. Ion dynamics, asymmetry, plasma jets. *Geophys. Res. Lett.* **51**, e2023GL105073 (2024).
55. E. Dubinin, R. Modolo, M. Fraenz, M. Päetzold, J. Woch, L. Chai, Y. Wei, J. E. P. Connerney, J. Mcfadden, G. DiBraccio, J. Espley, E. Grigorenko, L. Zelenyi, The induced magnetosphere of Mars: Asymmetrical topology of the magnetic field lines. *Geophys. Res. Lett.* **46**, 12722–12730 (2019).
56. M. Yan, J. Lei, N. Terada, B. Zhang, T. Dang, R. Sakata, S. Sakai, Y. Ma, Asymmetrically distributed Kelvin-Helmholtz instability at Venusian magnetosphere. *Geophys. Res. Lett.* **53**, e2026GL121691 (2026).
57. J. Luhmann, C. Russell, J. Spreiter, S. Stahara, Evidence for mass-loading of the Venus magnetosheath. *Adv. Space Res.* **5**, 307–311 (1985).
58. N. Terada, S. Machida, H. Shinagawa, Global hybrid simulation of the Kelvin–Helmholtz instability at the Venus ionopause. *J. Geophys. Res.* **107**, SMP 30–1–SMP 30–20 (2002).
59. T. Dang, J. Lei, B. Zhang, T. Zhang, Z. Yao, J. Lyon, X. Ma, S. Xiao, M. Yan, O. Brambles, K. Sorathia, V. Merkin, Oxygen ion escape at Venus associated with three-dimensional Kelvin-Helmholtz instability. *Geophys. Res. Lett.* **49**, e2021GL096961 (2022).
60. J. D. Huba, The Kelvin-Helmholtz instability: Finite Larmor radius magnetohydrodynamics. *Geophys. Res. Lett.* **23**, 2907–2910 (1996).
61. P. Henri, S. S. Cerri, F. Califano, F. Pegoraro, C. Rossi, M. Faganello, O. Šebek, P. M. Trávníček, P. Hellinger, J. T. Frederiksen, A. Nordlund, S. Markidis, R. Keppens, G. Lapenta, Nonlinear evolution of the magnetized Kelvin-Helmholtz instability: From fluid to kinetic modeling. *Phys. Plasmas* **20**, 102118 (2013).
62. J. Paral, R. Rankin, Dawn–dusk asymmetry in the Kelvin–Helmholtz instability at Mercury. *Nat. Commun.* **4**, 1645 (2013).
63. P. A. Delamere, R. J. Wilson, A. Masters, Kelvin-Helmholtz instability at Saturn's magnetopause: Hybrid simulations. *J. Geophys. Res.* **116**, A10222 (2011).
64. E. Liljeblad, T. Sundberg, T. Karlsson, A. Kullen, Statistical investigation of Kelvin-Helmholtz waves at the magnetopause of Mercury. *J. Geophys. Res. Space Phys.* **119**, 9670–9683 (2014).
65. H. Hasegawa, M. Fujimoto, K. Takagi, Y. Saito, T. Mukai, H. Rème, Single-spacecraft detection of rolled-up Kelvin-Helmholtz vortices at the flank magnetopause. *J. Geophys. Res.* **111**, A09203 (2006).
66. H. Hasegawa, A. Retinò, A. Vaivads, Y. Khotyaintsev, M. André, T. K. M. Nakamura, W. Teh, B. U. Ö. Sonnerup, S. J. Schwartz, Y. Seki, M. Fujimoto, Y. Saito, H. Rème, P. Canu, Kelvin-Helmholtz waves at the Earth's magnetopause: Multiscale development and associated reconnection. *J. Geophys. Res.* **114**, 2009JA014042 (2009).
67. T. K. M. Nakamura, M. Fujimoto, A. Otto, Structure of an MHD-scale Kelvin-Helmholtz vortex: Two-dimensional two-fluid simulations including finite electron inertial effects. *J. Geophys. Res.* **113**, A09204 (2008).
68. W. Li, M. André, Y. V. Khotyaintsev, A. Vaivads, D. B. Graham, S. Toledo-Redondo, C. Norgren, P. Henri, C. Wang, B. B. Tang, B. Lavraud, Y. Vernisse, D. L. Turner, J. Burch, R. Torbert, W. Magnes, C. T. Russell, J. B. Blake, B. Mauk, B. Giles, C. Pollock, J. Fennell, A. Jaynes, L. A. Avanov, J. C. Dorelli, D. J. Gershman, W. R. Paterson, Y. Saito, R. J. Strangeway, Kinetic evidence of magnetic reconnection due to Kelvin-Helmholtz waves. *Geophys. Res. Lett.* **43**, 5635–5643 (2016).
69. X. Ma, P. Delamere, A. Otto, B. Burkholder, Plasma transport driven by the three-dimensional Kelvin-Helmholtz instability. *J. Geophys. Res. Space Phys.* **122**, 10382–10395 (2017).

70. S. Ruhunusiri, J. S. Halekas, J. P. McFadden, J. E. P. Connerney, J. R. Espley, Y. Harada, R. Livi, K. Seki, C. Mazelle, D. Brain, T. Hara, G. A. DiBraccio, D. E. Larson, D. L. Mitchell, B. M. Jakosky, H. Hasegawa, MAVEN observations of partially developed Kelvin-Helmholtz vortices at Mars. *Geophys. Res. Lett.* **43**, 4763–4773 (2016).
71. D. Najib, A. F. Nagy, G. Tóth, Y. Ma, Three-dimensional, multifluid, high spatial resolution MHD model studies of the solar wind interaction with Mars. *J. Geophys. Res.* **116**, A05204 (2011).
72. J. P. McFadden, O. Kortmann, D. Curtis, G. Dalton, G. Johnson, R. Abiad, R. Sterling, K. Hatch, P. Berg, C. Tiu, D. Gordon, S. Heavner, M. Robinson, M. Marckwordt, R. Lin, B. Jakosky, MAVEN suprathermal and thermal ion composition (STATIC) instrument. *Space Sci. Rev.* **195**, 199–256 (2015).
73. J. E. P. Connerney, J. Espley, P. Lawton, S. Murphy, J. Odom, R. Oliversen, D. Sheppard, The MAVEN magnetic field investigation. *Space Sci. Rev.* **195**, 257–291 (2015).
74. G. Wang, S. Xiao, M. Wu, Y. Zhao, S. Jiang, Z. Pan, X. Hao, Y. Li, K. Liu, Y. Chi, Z. Zou, M. Chen, Z. Su, C. Shen, J. Guo, L. Cheng, Z. Wu, M. Xu, Y. Wang, T. Zhang, Calibration of the zero offset of the fluxgate magnetometer on board the Tianwen-1 orbiter in the Martian magnetosheath. *J. Geophys. Res. Space Phys.* **129**, e2023JA031757 (2024).
75. Y. Wang, T. Zhang, G. Wang, S. Xiao, Z. Zou, L. Cheng, Z. Pan, K. Liu, X. Hao, Y. Li, M. Chen, Z. Zhang, W. Yan, Z. Su, Z. Wu, C. Shen, Y. Chi, M. Xu, J. Guo, Y. Du, The Mars orbiter magnetometer of Tianwen-1: In-flight performance and first science results. *Earth Planet. Phys.* **7**, 216–228 (2023).
76. Z. Zou, Y. Wang, T. Zhang, G. Wang, S. Xiao, Z. Pan, Z. Zhang, W. Yan, Y. Du, Y. Chi, L. Cheng, Z. Wu, X. Hao, Y. Li, K. Liu, M. Chen, Z. Su, C. Shen, M. Xu, J. Guo, In-flight calibration of the magnetometer on the Mars orbiter of Tianwen-1. *Sci. China Technol. Sci.* **66**, 2396–2405 (2023).
77. J. W. Gao, Z. J. Rong, L. Klinger, X. Z. Li, D. Liu, Y. Wei, A spherical harmonic martian crustal magnetic field model combining data sets of MAVEN and MGS. *Earth Space Sci.* **8**, e2021EA001860 (2021).

**Acknowledgments:** We thank the MAVEN and Tianwen-1 instrument team for working in providing the data. We also appreciate the valuable discussions with J. McFadden and C. Fowler. Parts of this work for the observations obtained with MAVEN are supported by the French space agency CNES (National Centre for Space Studies). **Funding:** This work was supported by the NASA grant NNH10CC04C under the MAVEN project (to S.C.), the NASA Solar System Workings grants 80NSSC23K0911 (to C.D.) and 80NSSC24K1843 (to C.F.D.), the Alfred P. Sloan Research Fellowship (to C.D.), and the IBM Einstein Fellow Fund at the Institute for Advanced Study, Princeton (to C.D.). **Author contributions:** Conceptualization: C.Z., C.D., X.M., H.Z., J.H., and G.P. Methodology: C.Z., C.D., and J.H. Software: C.Z. Validation: H.Z. and K.G.H. Formal analysis: C.Z. and C.D. Investigation: C.Z., C.D., C.M., and H.Z. Resources: C.D. and S.C. Data curation: S.C., C.M., and K.G.H. Writing—original draft: C.Z. Writing—review and editing: R.S., X.L., C.D., C.M., X.M., H.Z., J.H., G.P., H.-W.S., S.C., L.W., and J.G. Visualization: C.Z. Supervision: C.D. and S.C. Project administration: C.Z., C.D., and S.C. Funding acquisition: C.D. **Competing interests:** The authors declare that they have no competing interests. **Data, code, and materials availability:** All data and code needed to evaluate and reproduce the conclusions in the paper are present in the paper and/or the Supplementary Materials. This study did not generate new materials. The MAVEN MAG data are publicly archived at https://pds-ppi.igpp.ucla.edu/mission/MAVEN/Magnetometer. The MAVEN STATIC data are publicly archived at https://pds-ppi.igpp.ucla.edu/mission/MAVEN/Supra-Thermal_and_Thermal_Ion_Composition. The Tianwen-1 MOMAG datasets are publicly available at http://space.ustc.edu.cn/dreams/tw1_momag/. The event list used in this study is publicly available via the Zenodo repository at https://doi.org/10.5281/zenodo.17354611. The corresponding source Python code for the analysis and figure generation can be accessed at https://doi.org/10.5281/zenodo.19635276.

Submitted 18 November 2025
Accepted 24 June 2026
Published 31 July 2026
10.1126/sciadv.aed9072

# Simultaneous Mars-orbit observations reveal Kelvin-Helmholtz instability–driven bulk atmospheric ion escape

Chi Zhang, Chuanfei Dong, Gangkai Poh, Jasper Halekas, Xuanye Ma, Ruhunusiri Suranga, Kathleen G. Hanley, Han-Wen Shen, Hongyang Zhou, Xinmin Li, Liang Wang, Jiawei Gao, Shannon Curry, and Christian Mazelle